\documentclass[aps,twocolumn,showpacs,amsmath,amssymb]{revtex4-1}

\usepackage{graphicx}
\usepackage{dcolumn}
\usepackage{bm}
\usepackage{verbatim}
\usepackage{hyperref}
\usepackage{float}
\usepackage{stackengine}

\graphicspath {{figures/}{../figures/}}

\begin{document}


\title{Setting nonperturbative uncertainties on \\ finite-temperature properties of neutron matter}

\author{Arianna Carbone}
\email[Email:~]{arianna.carbone@gmail.com}
\affiliation{European Centre for Theoretical Studies in Nuclear Physics and Related Areas (ECT*)
and Fondazione Bruno Kessler, Strada delle Tabarelle 286, I-38123 Villazzano (TN), Italy}

\date{\today}

\begin{abstract}
We present an error band on neutron matter properties at finite temperature (finite-T) which comprehends uncertainties on the nuclear interaction, the many-body method convergence, and the thermodynamical consistency of the approach. This study provides nonperturbative predictions for finite-T neutron matter employing chiral interactions which are selected on the basis of their performance in both finite nuclei and infinite matter at zero temperature. Since proper theoretical uncertainties at finite-T are still generally lacking, the band provided here represents a first step towards setting first-principles constraints on thermal aspects of the nuclear matter equation of state.
\end{abstract}

\maketitle


\section{Introduction}

In low-energy nuclear physics one is confronted with systems which size ranges from light isotopes (few \emph{fm}) to compact stars (tens of \emph{km}). The unifying relation between these apparently different realms, finite nuclei and neutron stars, is the nuclear matter equation of state (EoS)~\cite{Lattimer2012}. In this regime, the confinement properties of  quantum chromodynamics (QCD) are such that protons and neutrons, and possibly hyperons, are the common relevant degrees of freedom at play and, therefore, the knowledge of nuclear forces is key to understand the above phenomena. Deriving nuclear interactions directly from QCD is highly non trivial and nuclear physicists are forced to work the problem around by combining experimental information on finite nuclei, astrophysical observations and theoretical models. Consequently, constraints from different sources try to establish this structural equation for nuclear matter, among which are theoretical calculations of nuclear structure and reactions, systematics of nuclear masses and properties of isolated or merging neutron stars~(see Refs.~\cite{Oertel2017,Burgio2018} for recent reviews). 

\emph{Ab initio} nuclear theory plays a major role in this endeavor. For instance, both predictions of the neutron distribution in $^{48}$Ca~\cite{Hagen2015}, and calculations of neutron matter around saturation density~\cite{Hebeler2013Jul}, have put a window on the size of a neutron star, hence on its EoS, solely based on nuclear physics. Such first-principle predictions are further validated by the fact that the same nuclear forces are employed in finite nuclei and nuclear matter, where they have proved to work equally well~\cite{Ekstroem2015,Lapoux2016,Drischler2019,Idini:2019hkq,Holt:2019gmc,Leonhardt:2019fua}. Thanks to the first detected neutron-star merger event GW170817~\cite{Ligo2017}, outcomes of the analysis combined with neutron matter calculations~\cite{Most2018}, and also perturbative-QCD~\cite{Annala2018}, are putting challenging constraints on the radius of a neutron star. On top of this, the forthcoming measurements on the mass/radius of neutron stars via the X-ray telescope NICER could be valuable to help constrain both the low- and high-density regions of the nuclear EoS~\cite{Weih:2019rzo}. These new developments provide a clear path to nail down the correct relation between pressure and energy density of nuclear matter in the zero temperature (zero-T) limit. 

The situation is more uncertain when finite temperature (finite-T) is included in the EoS, condition which has to be considered when studying binary neutron star mergers or matter formed in heavy-ion collisions. Despite several first-principle calculations of nuclear matter at finite-T over the past forty years~\cite{Friedman1981,Haar1986,Lejeune1986,Huber1998,Baldo1999,Frick2003,Tolos2008,Soma2009,Rios2009,Fiorilla2012,Togashi2013,Wellenhofer2015,Carbone2018,Lu:2019mza}, there are only few attempts to include such results in EoSs for astrophysical simulations. Most applications have concentrated on the study of proto-neutron stars~\cite{Burgio2011,Chen2012,Camelio2017} or core-collapse supernovae~\cite{Togashi2017}. For binary neutron star mergers, the most widely used finite-T EoSs are usually based on Skyrme-like models, or on relativistic mean field theory~(for a full list of general purpose EoS see Refs.~\cite{Oertel2017,Burgio2018}).  
Another approach has been that of describing the thermal part of the EoS as an ideal fluid~\cite{Rezzolla_book:2013}. Nonetheless, this latter approximation can be quite crude for mimicking thermal properties of the EoS~\cite{Bauswein2010,Constantinou2015,Carbone2019,Lim:2019ozm}. Recently, finite-T EoSs for neutron stars have appeared based on quite different approaches~\cite{Du2019,Raithel:2019gws,Chesler:2019osn}. However, no detailed uncertainty analysis of nuclear matter properties that can constrain the finite-T EoS is available to date.

This work addresses this latter issue by performing a first-principles study of neutron matter at finite-T, that combines uncertainties on the nuclear interaction, the many-body approximation and the thermodynamical consistency of the approach. Since \emph{ab initio} results have shown to be quite constraining at nuclear densities to select acceptable zero-T EoSs from other approaches~\cite{Hebeler2013Jul}, this study wants to be a first step towards a similar selection based on nuclear physics including finite-T. Furthermore, we present a first complete analysis of the neutron effective mass at finite-T, quantity which has been shown to be crucial in determining the thermal behavior of the nuclear EoS~\cite{Constantinou2015,Yasin2018,Carbone2019,Schneider:2019shi}.

\section{Formalism}

Within a low-energy non-relativistic framework, we make use of the many-body self-consistent Green's function (SCGF) method~\cite{Barbieri2017} to investigate the properties of infinite nuclear matter employing chiral nuclear interactions~\cite{Epelbaum2009Oct,Machleidt2011,Hammer2013}. This approach takes into account beyond mean-field effects constructing a fully-dressed nucleon propagator~\cite{Dickhoff2004}. By solving the Dyson equation,
\begin{equation}
G({\bf p},\omega)=G_{0}({\bf p},\omega)+G_{0}({\bf p},\omega)\Sigma^\star({\bf p},\omega)G({\bf p},\omega)\,,
\label{eq:dyson}
\end{equation} 
a self-consistent description for the dressed propagator $G$ in terms of single-particle momentum ${\bf p}$ and energy $\omega$ is found, build upon its free version $G_0$. The nonperturbative self-energy $\Sigma^\star$ is constructed at each iterative step in the solution of the Dyson equation through a resummation of intermediate scattering diagrams, within the so called \emph{ladder approximation}~\cite{Frick2003,Rios2009,Soma2009}. Eq.~\eqref{eq:dyson} can be recast as a solution for the nucleon spectral function $\cal{A}$ given by the formula~\cite{Rios2009}:
\begin{equation}
{\cal A}({\bf p},\omega)=\frac{-2{\rm Im}\Sigma^\star({\bf p},\omega)}{\big[\omega-\frac{p^2}{2m}-{\rm Re}\Sigma^\star({\bf p},\omega)\big]^2+\big[\frac{-2{\rm Im}\Sigma^\star({\bf p},\omega)}{2}\big]^2}\,.
\label{eq:spectral_fun}
\end{equation}
A self-consistent solution is found when the spectral function which enters the calculation of the self-energy $\Sigma^\star$ equals the one obtained solving Eq.~\eqref{eq:spectral_fun}. With this spectral function it is possible to access the total energy per nucleon of the system via the Galitskii-Migdal-Koltun sumrule~\cite{Carbone2013}:
\begin{equation}
\frac{E}{A}=\frac{\nu}{n}\int\frac{{\rm d}{\bf p}}{(2\pi)^3}\int\frac{{\rm d}\omega}{2\pi}\frac{1}{2}\Big\{\frac{p^2}{2m}+\omega\Big\}{\cal A}({\bf p},\omega)f(\omega)-\frac{1}{2}\langle \hat W \rangle\,;
\label{eq:energy}
\end{equation}
$\nu$ is the degeneracy of the system, 2 for pure neutron matter (PNM) and 4 for symmetric nuclear matter (SNM); $n$ is the total number density; $\langle \hat W \rangle$ is the expectation value of the three-body operator. The SCGF method is implemented directly at finite-T and it provides in principle a thermodynamically consistent description of the many-body system, i.e. the microscopic and macroscopic (thermodynamical) estimates of physical properties should equal one another~\cite{Baym1961,Baym1962}. 

\section{Results}

\subsection{SNM, PNM and the symmetry energy at zero temperature with SCGFs}

\begin{figure}[t]
   \centering
        \includegraphics[width=0.48\textwidth]{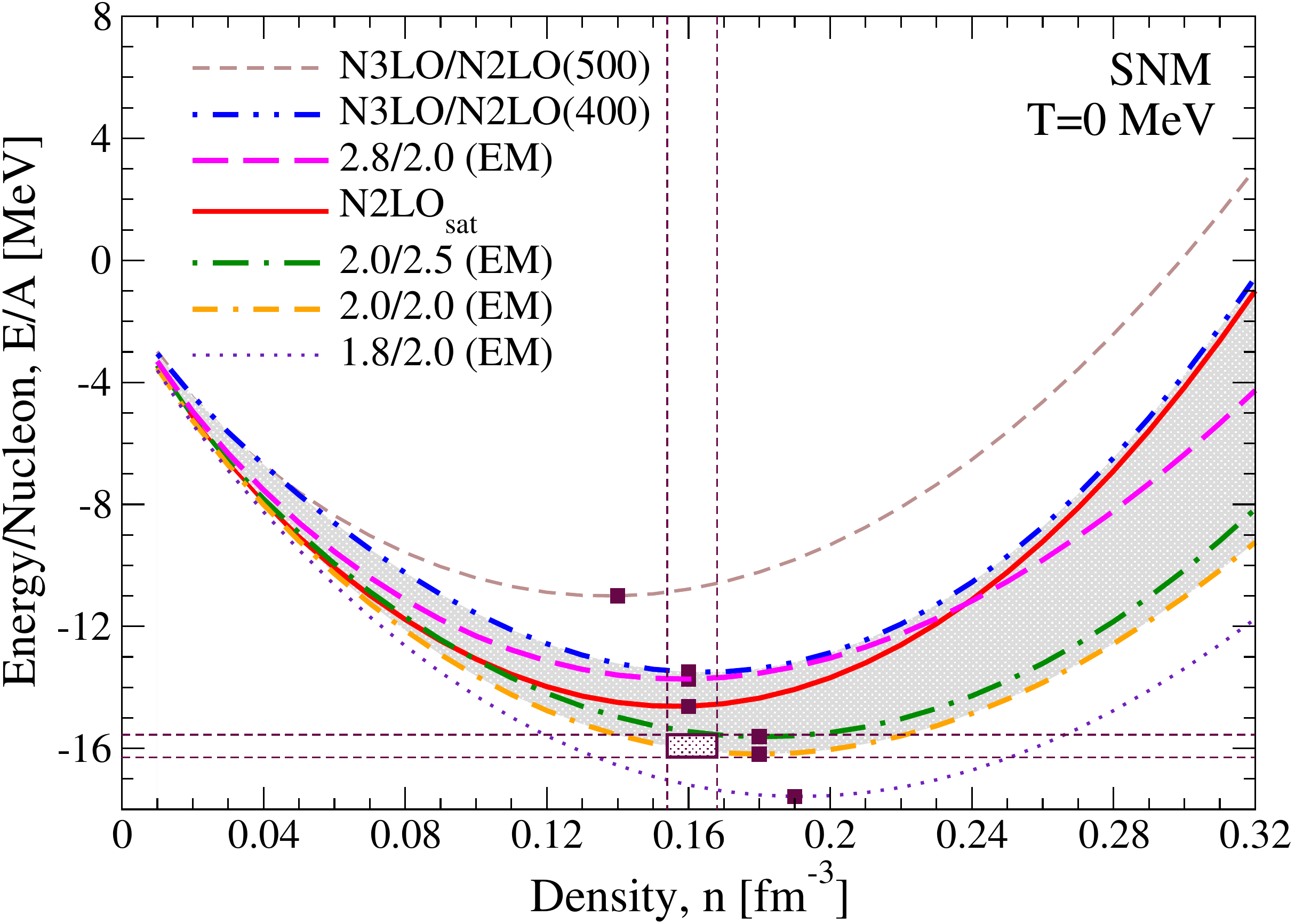}
    \caption{Zero-T SNM energy per nucleon vs. number density obtained within the SCGF method using seven chiral interactions (see text). Filled squares highlight the saturation point for each curve. Dotted box and relative dashed bands define the empirical saturation point as given by density functional theory~\cite{Brown2014}. Light/grey band is the accepted zero-T uncertainty band (see text for details).}
   \label{fig:ener_snm_T0}
\end{figure}

We start by analyzing in Fig.~\ref{fig:ener_snm_T0} the zero-T energy per nucleon of SNM obtained employing Eq.~\eqref{eq:energy}. We use  two plus three body chiral Hamiltonians which have proven successful in predicting finite nuclei properties: the N2LO$_{\rm sat}$ reproduces nuclear radii and binding energies up to $^{40}$Ca~\cite{Ekstroem2015,Lapoux2016}; the Entem-Machleidt (EM) potentials~\cite{Hebeler2011}, with labels $\lambda_{\rm 2N}/\Lambda_{\rm 3N}$ being the low-resolution scale on the two-body force and the cutoff on the three-body force, reproduce reasonably well the ground state energies of closed-shell nuclei, two-neutron separation energies and 2$^+$ excited states up to $^{78}$Ni~\cite{Simonis2017}; we have added two more interactions, dubbed N3LO/N2LO for the two-/three-body chiral force order, which, apart from being fit to properties of light nuclei as the previous potentials, are also fit to the triton beta decay, with label $(\Lambda)$ being the non-local cutoff on the current~\cite{Klos2016}. Details for each Hamiltonian can be found in the corresponding Refs.~\cite{Ekstroem2015},\cite{Hebeler2011},\cite{Klos2016}.
We obtain a spread of $\sim$7 MeV in energy range at saturation density $n_{\rm sat}=0.16$fm$^{-3}$. However predicted saturation points build a Coester-like line, as also seen in Ref.~\cite{Drischler2019}. We use the empirical saturation box in Fig.~\ref{fig:ener_snm_T0} to select five interactions out of the original seven, based on being the predicted saturation point consistent with either the density or energy ranges of the box (a light/grey band highlights the chosen interactions). The box in Fig.~\ref{fig:ener_snm_T0} is given by 12 Skyrme functional calculations, constrained by properties of doubly magic nuclei and \emph{ab initio} low-density neutron matter~\cite{Brown2014}. The predicted saturation points lie then in a range of densities $n$=[0.16-0.18]fm$^{-3}$ and energies $E/A$=-[13.5-16.2]MeV.

\begin{figure}[t]
\includegraphics[width=.48\textwidth]{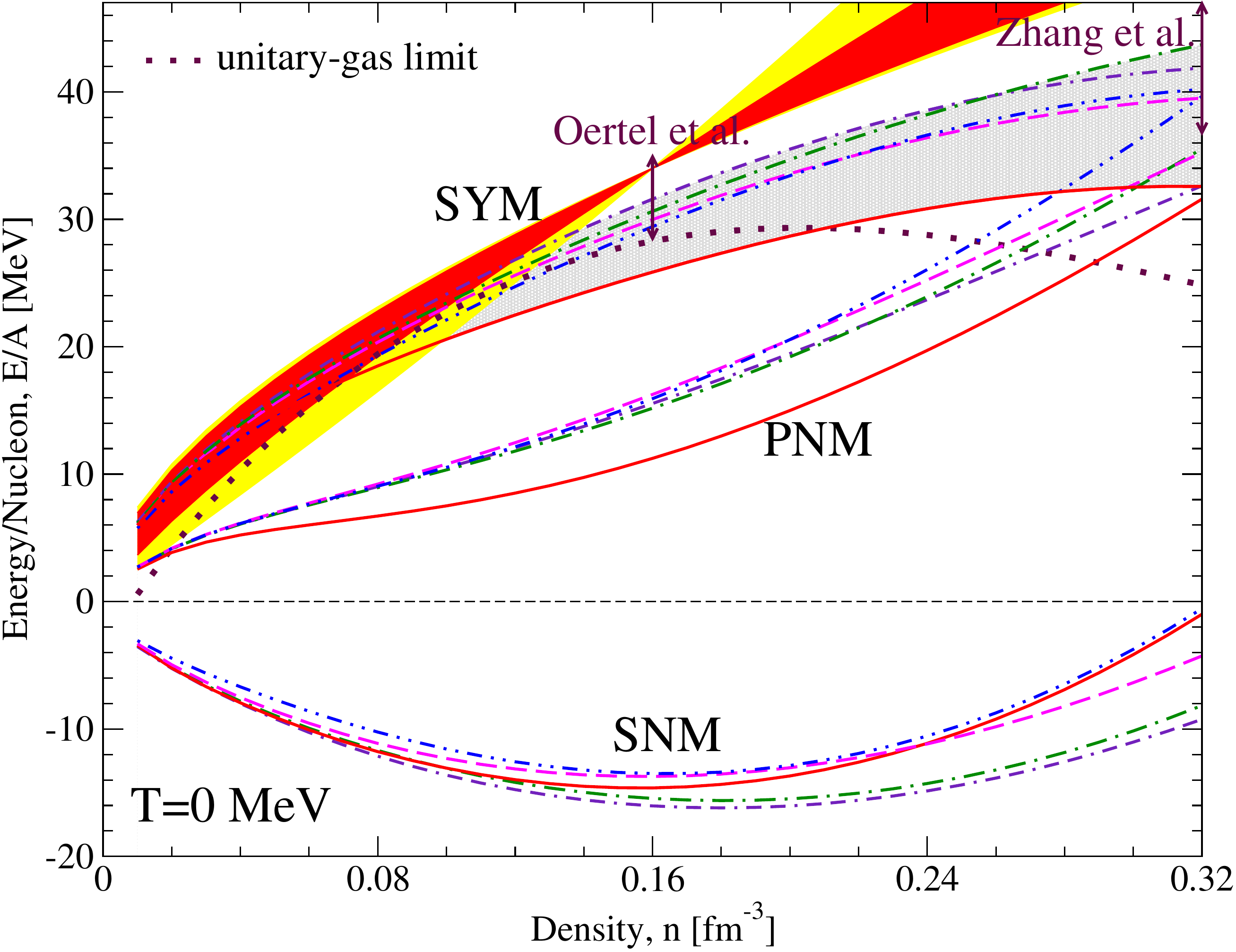}
\caption{Zero-T SNM, PNM, and symmetry (SYM) energies (highlighted with a band) vs. number density obtained within the SCGF method for five selected chiral interactions, legend follows that in Fig.~\ref{fig:ener_snm_T0}. Dotted line is the conservative unitary-gas limit given by Ref.~\cite{Tews2017}. Intervals at $n_{\rm sat}$ and 2$n_{\rm sat}$ come from Ref.~\cite{Oertel2017} and \cite{Zhang2019}. Lighter/yellow and darker/red bands define respectively the FOPI/LAND and ASY-EOS results extracted from reaction experiments~\cite{Russotto2016}.}
\label{fig:symm_ener}
\end{figure}

The selected interactions are used in Fig.~\ref{fig:symm_ener} to calculate the symmetry energy SYM (highlighted with a band), which is obtained as the difference between the energies of PNM and SNM, also shown in the figure. All SYM curves stand together except for the N2LO$_{\rm sat}$ calculation. This is caused by an extremely soft PNM energy per nucleon. We delimit the predicted symmetry energies from below exploiting the unitary-gas limit (dotted line) corresponding to the conservative choice of Ref.~\cite{Tews2017}. While all calculations respect this lower bound, the N2LO$_{\rm sat}$ violates it in a region of densities from $n_{\rm sat}/2$ to $1.25n_{\rm sat}$. Furthermore, it is the only case which does not match the comprehensive uncertainty interval given by Oertel \emph{et al.}~\cite{Oertel2017}, which constrains the symmetry energy at $n_{\rm sat}$ employing a large number of theoretical and experimental calculations from different sources. The uncertainty interval at $2n_{\rm sat}$ provided by Zhang \emph{et al.}~\cite{Zhang2019}, matched by all calculations except for N2LO$_{\rm sat}$, is extracted constraining an explicitly isospin-dependent parametric EoS with the data analysis of GW170817~\cite{Ligo2018}. It is interesting to note that, except for a region with densities below $\sim$0.12fm$^{-3}$, it is not possible to reconcile our theoretical predictions with the symmetry energy extracted from the measured $^{197}$Au+$^{197}$Au reaction data from the FOPI-LAND (lighter/yellow band in Fig.~\ref{fig:symm_ener}) and ASY-EOS (darker/red band) experiments~\cite{Russotto2016}. The experimental extracted value of 34MeV at $n_{\rm sat}$ stands higher than the entire uncertainty band we provide, with growing departure as density increases~\cite{Russotto2016}. The explicit inclusion of the $\Delta$ degree of freedom could help improve this issue~\cite{Ekstroem2018}.

\begin{figure}[t]
\includegraphics[width=.48\textwidth]{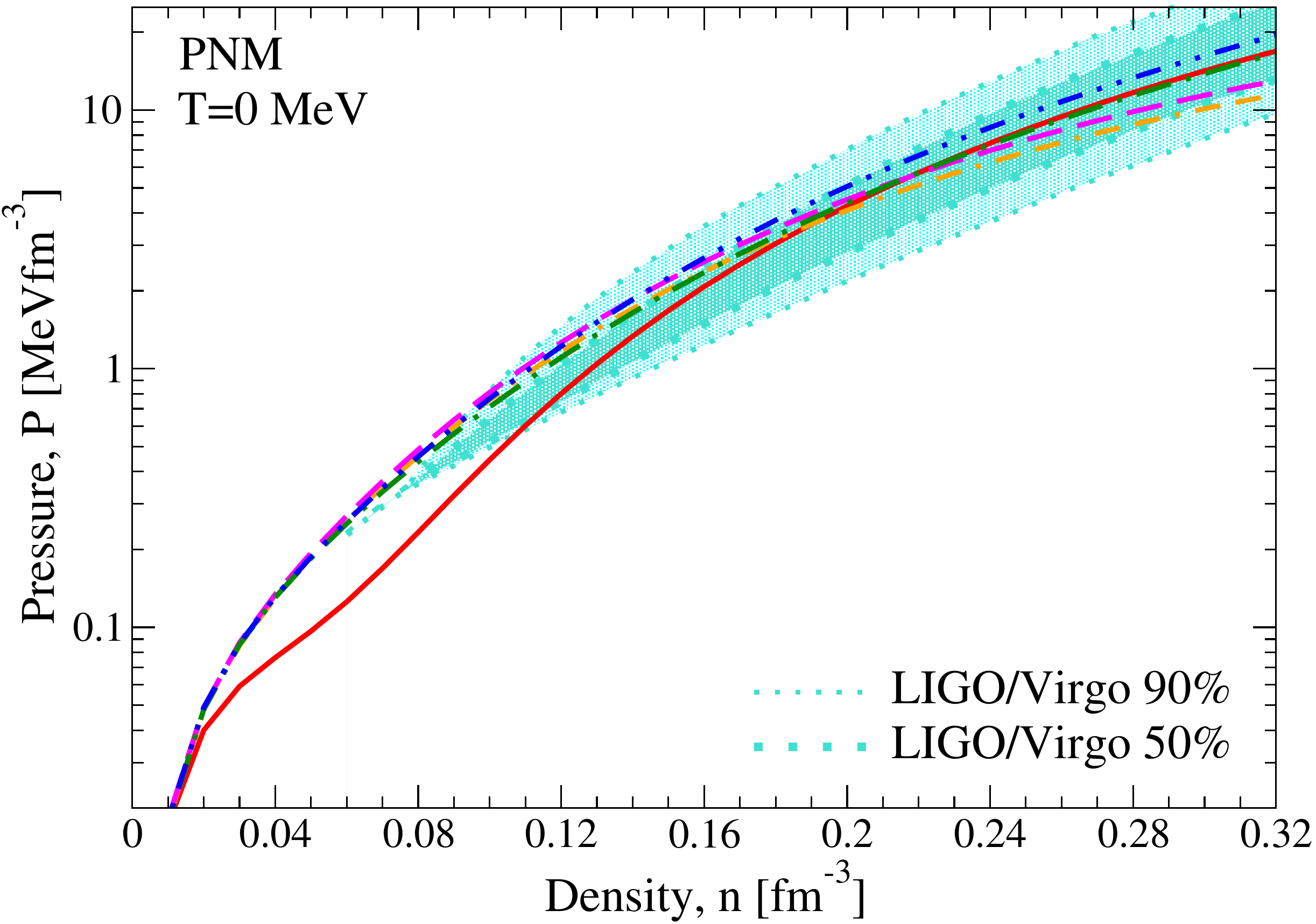}
\caption{Zero-T PNM pressure vs. number density obtained within the SCGF method employing the same chiral interactions of Fig.~\ref{fig:symm_ener}, legend follows that in Fig.~\ref{fig:ener_snm_T0}. Small and big dotted lines delimit the LIGO/Virgo 90\% and 50\% confidence level bands on the neutron star matter pressure as obtained from the analysis of GW170817~\cite{Ligo2018}.} 
\label{fig:press_t0}
\end{figure}

In spite of the apparently poor performance of N2LO$_{\rm sat}$, we choose to use the entire band of the five interactions to calculate the zero-T pressure in PNM. This choice is also based on the outstanding performance of N2LO$_{\rm sat}$ in finite nuclei~\cite{Hagen2015,Lapoux2016,Idini:2019hkq}. Figure~\ref{fig:press_t0} compares our results with the constrained bands at a 90$\%$ and 50$\%$ confidence level as given by the analysis of GW170817~\cite{Ligo2018}. All calculations are very close to the LIGO/Virgo bands at low densities (results from Ref.~\cite{Ligo2018} start at $\sim n$=0.06fm$^{-3}$), except for the N2LO$_{\rm sat}$ which remains very soft. However, as it appears clearly from Fig.~\ref{fig:press_t0}, the N2LO$_{\rm sat}$ pressure grows fast approaching the other calculations around $n_{\rm sat}$. The theoretical calculations then stay within the 50$\%$ confidence level band all the way up to $\sim2n_{\rm sat}$. This comparison underlines the fact that, even though the PNM energy predicted could be low, as in N2LO$_{\rm sat}$, its derivative to obtain the pressure can be in any case quite steep. For a proper comparison with the LIGO/Virgo bands one should consider the presence of protons; a fraction up to $\sim10\%$ would lower the theoretical pressure by $\sim15\%$ of its value~\cite{Wellenhofer2015}, which overall maintains the validity of the discussion presented for Fig.~\ref{fig:press_t0}.

Based on all of the above considerations, we select two final chiral interactions to generate an uncertainty band to study finite-T properties of neutron matter: the upper limit given by the N2LO$_{\rm sat}$ and lower limit by the 2.0/2.0(EM). We restrict the upper limit of the grey band in Fig.\ref{fig:ener_snm_T0} to N2LO$_{\rm sat}$ on the basis of the closest predictions to the SNM empirical saturation point, and given that the error at $2n_{\rm sat}$ encompasses practically all the five previously selected ones, so we are sure to keep the uncertainty at high densities as conservative as possible (this applies also for symmetry energy and pressure in Figs.~\ref{fig:symm_ener}-\ref{fig:press_t0}). Furthermore, our choice is corroborated by a reliable prediction of the SNM liquid-gas phase transition obtained using these interactions~\cite{Carbone2018}. 

\begin{figure*}[t]
   \centering
        \includegraphics[width=0.48\textwidth]{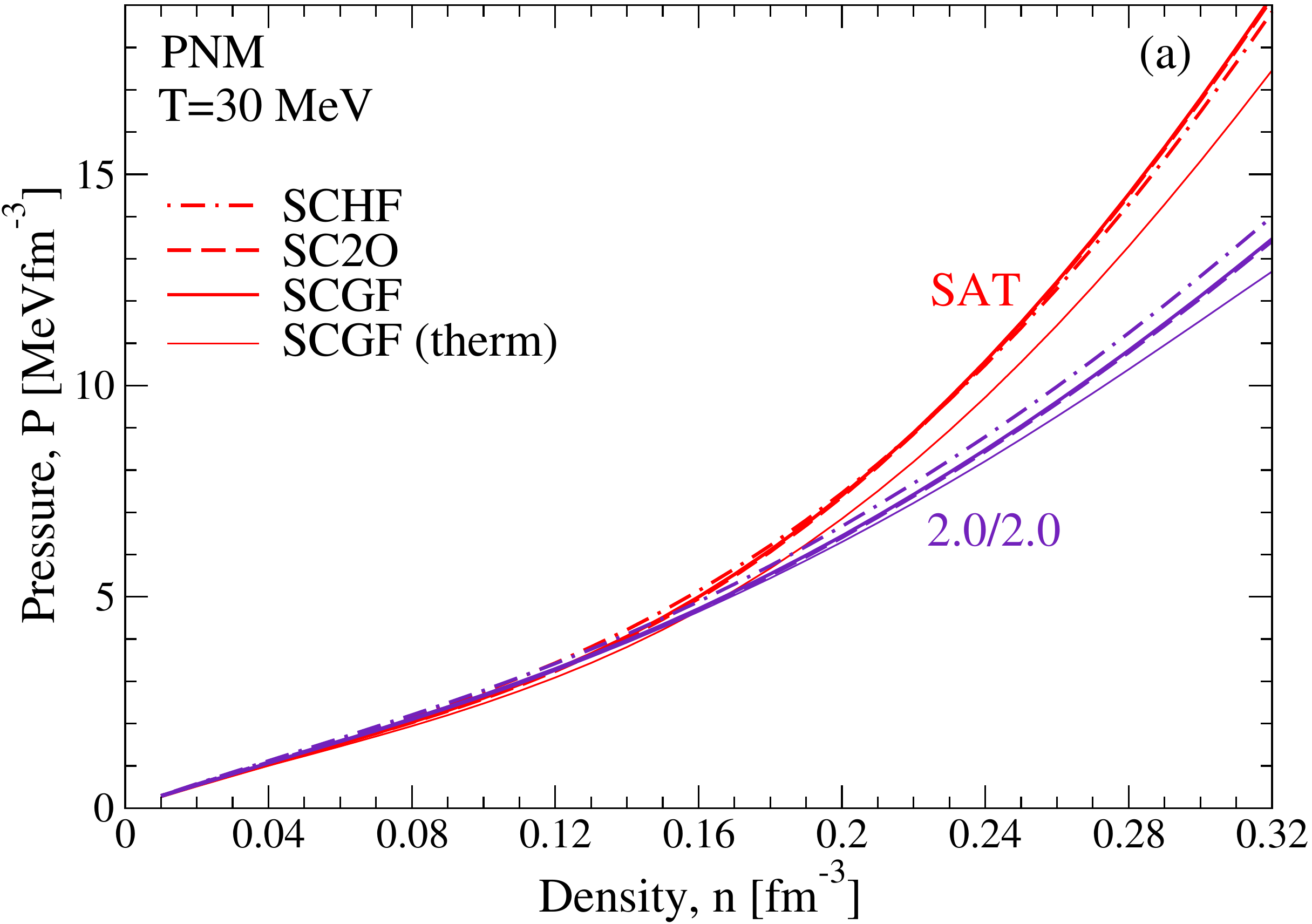}
        ~
        \includegraphics[width=0.48\textwidth]{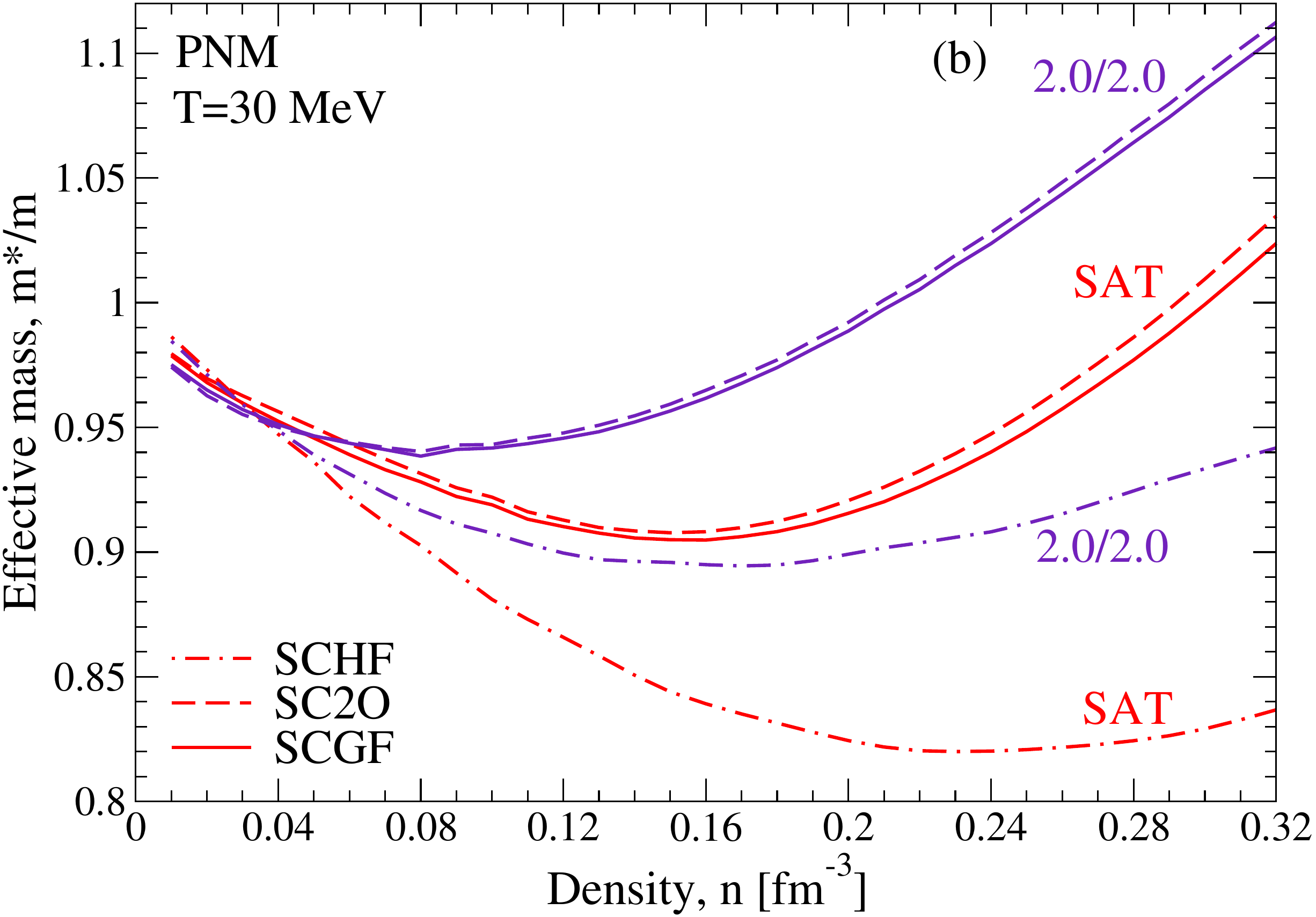}

        \includegraphics[width=0.48\textwidth]{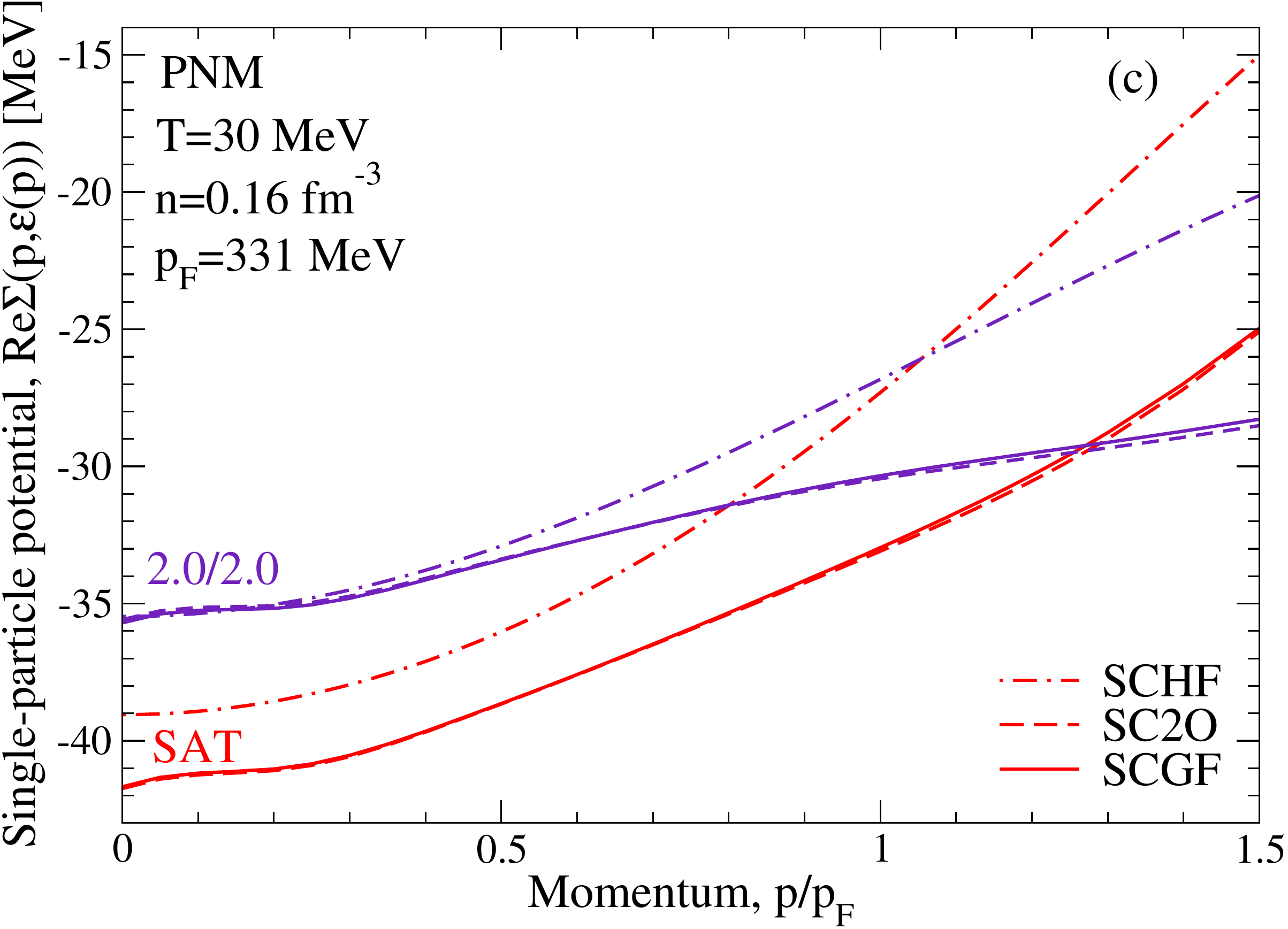}
        ~
        \def\stackalignment{l}\topinset{\includegraphics[width=0.18\textwidth]{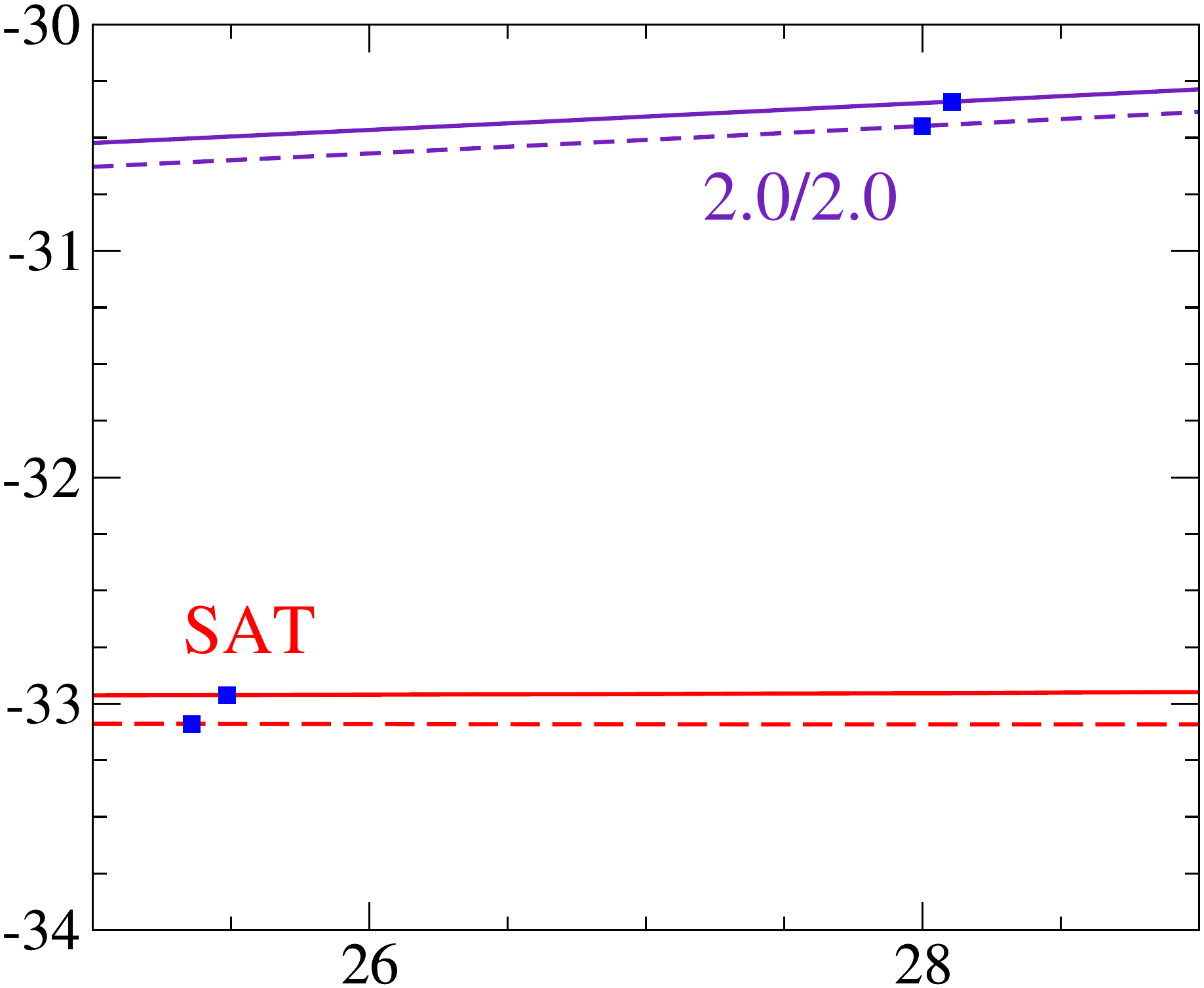}}{\includegraphics[width=0.48\textwidth]{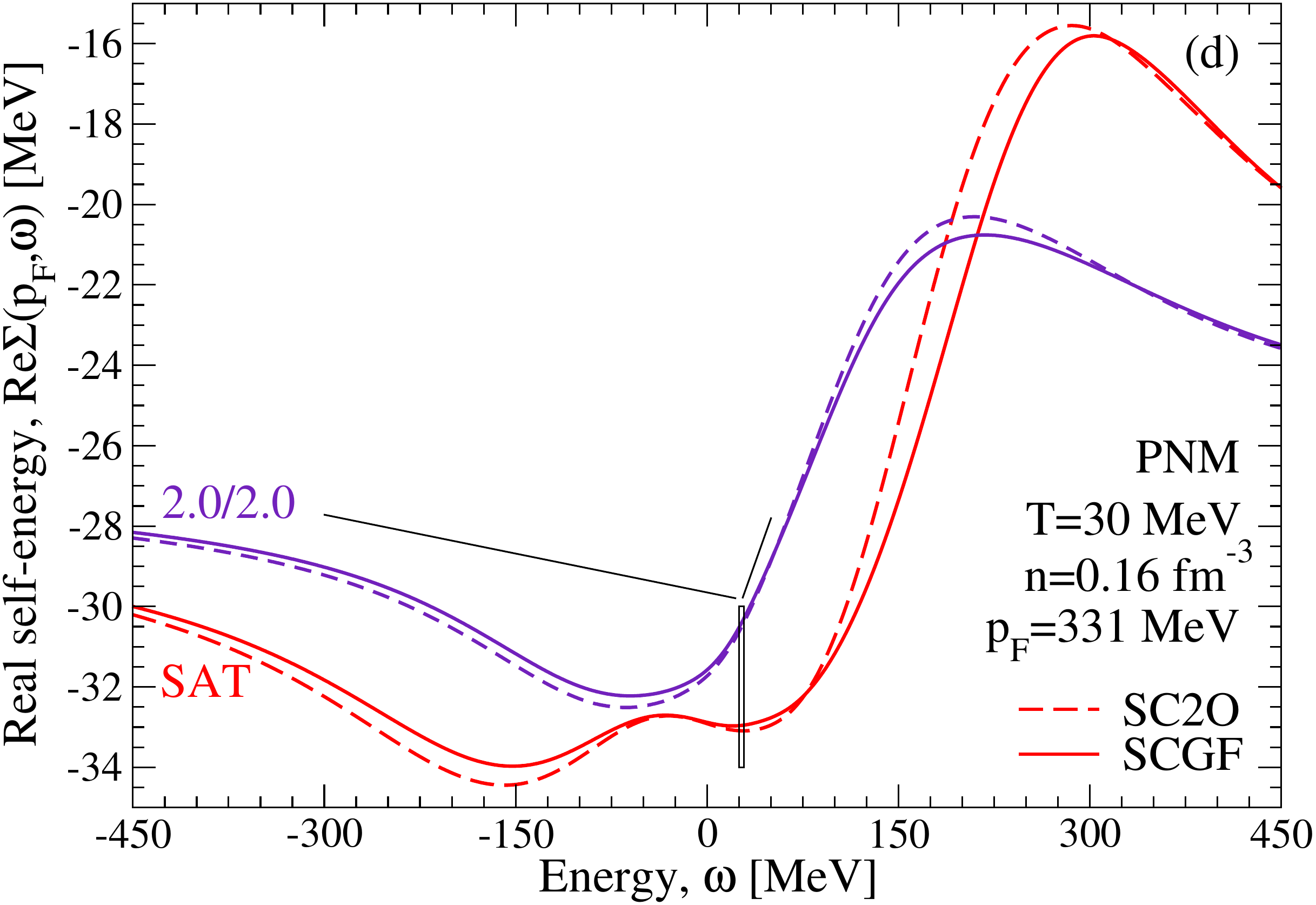}}{15pt}{55pt}

         \caption{PNM properties at T=30 MeV obtained using different approximations of the many-body method: SCHF, SC2O and SCGF (see text for details), employing two of the interactions of Fig.~\ref{fig:ener_snm_T0}, label SAT for N2LO$_{\rm sat}$ and label 2.0/2.0 for the 2.0/2.0 (EM). Panel (a): pressure vs. number density. Panel (b): neutron effective mass vs. number density. Panel (c): single-particle potential as function of momentum at $n_{\rm sat}$. Panel (d): real self-energy as function of energy at $n_{\rm sat}$ for $p=p_{\rm F}$; the inset shows the points (blue dots) where the derivative of the self-energy is performed to obtain the $m^\omega$ (see text for details).}
    \label{fig:finiteT_prop}
\end{figure*}

We must point out that proper uncertainty quantification should be performed order by order in the chiral effective field theory expansion, with cutoff variations and pertinent refit of low-energy constants at each order, as proposed for example in Refs.~\cite{Epelbaum2015,Carlsson2016}. However these Hamiltonians have not yet proven to be successful to reproduce the properties of $^{16}$O or heavier nuclei~\cite{Epelbaum2019,Carlsson2016}, nor nuclear matter~\cite{Drischler2019}, and for such we stick at present to the above mentioned chiral interactions. The latter rely nonetheless on a power counting scheme affected by non renormalization-group invariance, hindering a proper connection with QCD~\cite{Hammer:2019poc}. The use of alternative chiral models exiles the scope of the present paper, since it is not yet foreseeable when these interactions will become available for \emph{ab initio} calculations. To enhance furthermore the connection to QCD, fitting the low-energy constants to LatticeQCD calculations would be the desirable way to go, as presented recently for the nucleon axial coupling in Ref.~\cite{Chang:2018uxx}. Until the above will be addressed, the predictions of the present available interactions will bear some model dependence to the specific data they are fit to, but they remain the best option to maximize a link to QCD.

\subsection{PNM finite-temperature properties with SCGFs}

We present in Fig.~\ref{fig:finiteT_prop} specific PNM properties at T=30 MeV, chosen as a representative temperature, employing the above mentioned chiral interactions. In panel (a) we show the pressure as a function of number density. To understand the uncertainty we have on the many-body calculation and test the method convergence, we present three different approximations: self-consistent Hartree-Fock (SCHF), self-consistent second order (SC2O) and SCGF, where the self-energy is truncated respectively at first-, second- and all-orders (\emph{ladder}). Furthermore, we test the thermodynamical consistency of our calculations by comparing the SCGF pressure obtained via the microscopic chemical potential $\tilde\mu$, i.e. $\tilde P=n(\tilde\mu-F/A)$, with respect to the pressure obtained thermodynamically via derivative of the free-energy per nucleon $F/A$, i.e. $P=n^2\frac{\partial F/A}{\partial n}$, dubbed SCGF (therm) (see Ref.~\cite{Carbone2018} for further details). 
On one hand we see how, for both chiral interactions, the SC2O calculations stand practically on top of the full SCGF ones, meaning that self-energy terms beyond second-order are quite small for these interactions. On the other hand, the differences arising between the microscopic and macroscopic derivation of the pressure, SCGF vs SCGF(therm), are more visible. Thermodynamical consistency holds when only two-body forces are considered~\cite{Rios2009}, so this issue is related to the inclusion of three-body forces and specifically to the approximation with which we calculate $\langle\hat W\rangle$ in Eq.~\eqref{eq:energy}~\cite{Carbone2014,Carbone2018}. It is instructive to see how even the SCHF first order calculations stand very close to the full SCGF ones, providing at $2n_{\rm sat}$ a band of $\sim$1.5MeV for the full many-body method uncertainty, compared to $\sim$6 MeV coming from the chiral interaction error band. It must be noted that at twice saturation density 
one is probing the range of the resolution scale of the chiral Hamiltonians employed, thus approaching the limit of validity of these interactions. We account for such uncertainty by performing calculations with different chiral interactions.

In Ref.~\cite{Carbone2019} we provided the thermal indexes, i.e. a quantity which characterizes the finite-T EoS,  obtained from the chiral interactions employed in Fig.~\ref{fig:finiteT_prop}. Compared to the tests performed in Ref.~\cite{Bauswein2010}, the uncertainty in the present results is tighter and could help better locate the main gravitational-wave frequency peak of the postmerger remnant (see also Ref.~\cite{Baiotti2017} for tests using a number of different finite-T EoSs). Certainly, to precisely assess the impact of such theoretical calculations, one would have to extend these to the needed higher densities and perform the full general-relativity simulation.

Panel (b) of Fig.~\ref{fig:finiteT_prop} displays the neutron effective mass at T=30 MeV calculated at the Fermi momentum $p=p_{\rm F}$ and Fermi energy $\omega=\varepsilon(p_F)$ for each density. Contrary to the pressure, a big discrepancy is visible here between the SCHF and the SC2O/SCGF calculations. Apart from the differences in the self-energy truncations, this behavior is to be ascribed also to the fact that at the SCHF level we only have one term for the effective mass, $m^*/m=m^k=\left(1+\frac{m}{p}\frac{\partial{\rm Re}\Sigma(p,\varepsilon(p))}{\partial p}\right)^{-1}$, while for both SC2O and SCGF we also have a contribution from the $m^{\omega}=1-\frac{\partial{\rm Re}\Sigma(p,\omega)}{\partial\omega}$, which leads to a total effective mass of $m^*/m=m^km^\omega$. These two quantities, $m^k$ and $m^\omega$,  incorporate the properties of the varying single-nucleon self-energy in terms of momentum and energy respectively, measuring its non locality either in space or time~\cite{BALi2018}.  Similar to the pressure in panel (a), the SC2O $m^*$ is already a good approximation of the full SCGF results. The rising of the effective mass with density after reaching a certain minima is caused by the inclusion of three-body forces~\cite{Hebeler2010,Arellano2016,Burac2019}. 

To understand more in depth the behavior of the effective mass, we present in panel (c) and (d) of Fig.~\ref{fig:finiteT_prop} the single-particle potential as a function of momentum and the real self-energy as a function of energy, which contribute to the calculation of the $m^k$ and $m^\omega$ respectively. The single-particle potential in panel (c) corresponds to an ``on-shell" real self-energy, ${\rm Re}\Sigma(p,\varepsilon(p))$, obtained solving a self-consistent equation for the single-particle energy~\cite{Rios2009,Xu2019}. A more repulsive single-particle potential and a steeper behavior for the SCHF case is what causes the $m^k$ to become smaller, and consequently obtain a smaller effective mass with respect to the SC2O/SCGF cases (see panel (b)). 
In panel (d) of Fig.~\ref{fig:finiteT_prop} we plot the self-energy as a function of energy calculated at $p_{\rm F}$ for $n_{\rm sat}$.  The blue dots in the inset show the points, $\omega=\varepsilon(p_F)$, where the derivative of the self-energy is performed for the calculation of the $m^\omega$, highlighting quite a different steepness for the two interactions. The inclusion of the $m^\omega$ is found to be fundamental to reproduce the behavior of thermal effects in finite-T EoSs~\cite{Carbone2019}. In fact, knowledge of the effective mass provides a further independent evaluation of the thermal index without the need to calculate bulk properties of the many-body system~\cite{Carbone2019}. This could help to further assess the graviational-wave spectrum of the postmerger phase~\cite{Bauswein2010}.

\section{Conclusion}

To conclude, we employed state-of-the-art chiral interactions, proven successful in describing finite nuclei and zero-T infinite matter, to set an uncertainty band on finite-T properties of neutron matter. We provided an error analysis that accounts for uncertainties in the interaction, the many-body method truncation and the thermodynamical consistency of the approach. The major uncertainty on the PNM finite-T pressure is related to the chiral interaction, being this at $2n_{\rm sat}$ four times larger than the error associated to the many-body method. On the contrary, this behavior is reversed for microscopic properties, such as the effective mass or the single-particle potential, where the full many-body error is more than twice the chiral interaction uncertainty band at $2n_{\rm sat}$. This underlines the fact that beyond first-order calculations of the nucleon effective mass are mandatory, especially in view of recent studies which show how parametrized functions of this quantity can mimick the thermal part of the EoS~\cite{Carbone2019}. Furthermore, the value of the effective mass has been shown to be crucial for the onset of explosion in core-collapse supernovae~\cite{Yasin2018,Schneider:2019shi}. A direct improvement of this study, towards the determination of a global uncertainty band which combines together all the sources of error here enumerated, would be to perform Bayesian analysis on the properties calculated at finite-T (see for example Refs.~\cite{Furnstahl2015,Lim2018}). One would employ the latter results as priors to predict the most probable values of such properties with their confidence intervals. This study of nonperturbative uncertainties on finite-T properties of neutron matter then represents a first fundamental step to reliably constrain the nuclear EoS employed in astrophysical simulations of merging neutron stars~\cite{Bauswein2010,Bauswein:2019ybt,Koppel2019,Perego2019,Kiuchi2019}.\\

\begin{acknowledgments}
The Author is grateful to Carlo Barbieri for a thorough reading of the manuscript and for much needed comments. Fundamental discussions with Arnau Rios on the numerical calculations are acknowledged. Calculations for this research were conducted on the Lichtenberg high-performance computer of the TU Darmstadt.
\end{acknowledgments}

\bibliographystyle{apsrev4-1}
\bibliography{biblio}

\end{document}